\documentclass[pra,10pt,twocolumn,superscriptaddress,floatfix,showpacs]{revtex4-1}

\usepackage{subfig}
\usepackage{graphicx} 
\usepackage{epstopdf}
\usepackage[utf8]{inputenc}
\usepackage[T1]{fontenc}     
\usepackage[british]{babel}  
\usepackage[sc,osf]{mathpazo}
\usepackage[scaled=0.86]{berasans}  
\usepackage[colorlinks=true, citecolor=blue, urlcolor=blue]{hyperref}  
\usepackage[babel]{microtype}  
\usepackage{amsmath,amssymb,amsthm,bm,amsfonts,mathrsfs,bbm} 

\usepackage{xspace}  
\usepackage{pgfplots}
\usepackage{xcolor,colortbl}
\def\ba{\begin{equation}}
\def\ea{\end{equation}}
\def\bea{\begin{eqnarray}}
\def\eea{\end{eqnarray}}
\def\ben{\begin{equation*}}
\def\een{\end{equation*}}
\def\bean{\begin{eqnarray*}}
\def\eean{\end{eqnarray*}}
\def\bma{\begin{mathletters}}
\def\ema{\end{mathletters}}
\def\bi{\begin{itemize}}
\def\ei{\end{itemize}}

\newcommand{\be}{\begin{equation}}
\newcommand{\ee}{\end{equation}}

\newcommand{\kommentar}[1]{}

\newcommand{\forget}[1]{}

\begin{document}

\title{Characterizing Quantum Correlations In Fixed Input $n$-Local Network Scenario }

\author{Kaushiki Mukherjee}
\email{kaushiki_mukherjee@rediffmail.com}
\affiliation{Department of Mathematics, Government Girls' General Degree College, Ekbalpore, Kolkata-700023, India.}
\author{Biswajit Paul}
\email{biswajitpaul4@gmail.com}
\affiliation{Department of Mathematics, Balagarh Bijoykrishna Mahavidyalaya, Balagarh, Dist. - Hooghly-712501, India}
\author{Arup Roy}
\email{arup145.roy@gmail.com}
\affiliation{S. N. Bose National Centre for Basic Sciences, Salt Lake, Kolkata 700 106, India}


\begin{abstract}
Contrary to Bell scenario, quantum nonlocality can be exploited even when all the parties do not have freedom to select inputs randomly. Such manifestation of nonlocality is possible in networks involving independent sources. One can utilize such a feature of quantum networks for purpose of entanglement detection of bipartite quantum states. In this context, we characterize correlations simulated in networks involving finite number of sources generating quantum states when some parties perform fixed measurement. Beyond bipartite entanglement, we enquire the same for networks involving sources now generating pure tripartite quantum states. Interestingly, here also randomness in input selection is not necessary for every party to generate nonlocal correlation.
\end{abstract}

\maketitle

	
\section{Introduction}
Entanglement of multipartite quantum system \cite{horodecki11} plays a pragmatic role in manifesting deviation of Quantum Theory (QT) from classical world. J.T.Bell used this intrinsic feature of the theory to abandon the possibility of existence of any local realistic interpretation of QT \cite{RRR,RLL} which however respects no-signaling principle. Bell's Theorem provides empirical methodology to detect nonlocal behavior of quantum correlations (often referred to as Bell nonlocality), experimental demonstration of which have already been provided \cite{as1,as}. \\
Speaking of tests demonstrating Bell nonlocality, the most simplest test was proposed by Clauser \emph{et al.} \cite{Cl}. Such a test involves two distant observers (Alice and Bob,say) such that each of them performs one binary measurement choosing randomly from a set of two measurements. To be precise, Alice randomly chooses one input from set of two inputs ($\{\mathcal{A}_0,\mathcal{A}_1\}$,say) and similarly Bob randomly chooses input from another set, say $\{\mathcal{B}_0,\mathcal{B}_1\}.$ Moreover, choice of inputs of Alice does not depend on that of Bob and vice-versa (measurement independence)$.\,\textmd{Bipartite}$ correlations generated after measurements are used in testing correlators based inequality, more commonly referred to as CHSH inequality$.\,$Violation of CHSH inequality indicates nonlocal nature of corresponding correlations. Till date, analogous to CHSH inequality \cite{Cl}, different correlators based inequalities (referred to as Bell inequalities) have been derived. Detecting quantum nonlocality by any of these tests requires randomness in choice of inputs of both the observers present in the corresponding measurement scenario. However, random selection of inputs by all observers is not a necessity to exploit non classicality of quantum correlations simulated in network scenarios \cite{ng1,ng2,ng3,ng4,ng5,ng6} characterized by source independence (often referred to as $n$-local networks).\\
 $\,\,$$n$-local quantum networks \cite{BRAN,BRA,star,km1,km2,km3,km4} basically refers to a network of $n$ sources, independent of each other, such that each of these sources generates an $m$-partite quantum state ($m$$\geq$$ 2$) shared between $m$ distinct parties. Nonlocality of correlations generated in such networks was first observed in bilocal ($n,m$$=$$2$) network \cite{BRAN,BRA} where entanglement was distributed from two independent sources. Such type of nonlocality is referred to as non bilocality \cite{BRAN,BRA}. Non bilocality, or in more general non $n$-locality  differs from usual sense of Bell nonlocality (standard nonlocality) where entanglement is distributed from a common source. Some of the measurement scenarios involved in $n$-local networks have been proposed where some ($P_{14}$ or $P_{13}$ measurement scenarios \cite{BRAN}) or all \cite{ng2,ng3} of the observers perform single measurement (referred to as `fixed measurement' \cite{ng3})$.\,\,$All of these studies basically analyzed some specific instances of quantum non $n$-locality in such measurement scenarios where not all observers \cite{BRAN} can randomly select inputs thereby manifesting instances of `quantum nonlocality without inputs' \cite{ng3}. Now observation of quantum nonlocality in network can be used for the purpose of detection of quantum entanglement in the same. In this context, we first intend to exploit quantum nonlocality in networks, characterized by source independence and fixed input criterion (for at least one of the observers). Quantum networks witnessing non $n$-locality can then be used for detection of entanglement resources.\\
For our purpose, we first characterize quantum correlations thereby analyzing non $n$-local nature of the correlations as detected via violation of existing non $n$-local inequality \cite{km1} when each of the sources generates an arbitrary two qubit state. In this context, one may note that such a study of quantum violation was recently initiated in \cite{ng1} where only two entangled sources ($n$$=$$2$) were considered (bilocal network).\\
As a direct consequence of our findings in practical ground, we propose a scheme of detecting entanglement (if any) using networks involving independent sources where not all parties have access to random choice of inputs. Such a protocol, relying on $n$-local correlations generated in `fixed input' measurement scenario \cite{km1}, serves the purpose of bipartite entanglement detection. Fixing measurements of some of the parties makes implementation of our protocol easier compared to the simplest standard Bell scenario of measurements$.\,$However, it must be pointed out that more easier implementation of protocols (compared to the scenario to be considered presently) may be possible if parties are allowed to randomly select from some suitable measurement settings which are more easily implementable. But we do not consider those easily implementable measurement setting scenarios with a motivation to detect entanglement in absence of random input selections.  \\
Recently $n$-local networks involving sources distributing multipartite entanglement have been designed in \cite{km2}$.\,\,$However, in such measurement scenario, all the parties had access to random choice of measurements. To verify quantum nonlocality even in absence of randomness in input selection (by some of the parties), we consider measurement scenario where now three independent sources generate tripartite quantum states. In this context, we have designed a set of non-linear Bell inequalities, violation of which suffice to detect non $n$-locality. The non-linear trilocal network scenario is then used for the purpose of tripartite entanglement detection. Our protocol (characterized by fixed measurement setting by two parties) can detect both biseparable and genuine entanglement (some members of GGHZ and W classes) of pure  tripartite quantum states. Interestingly it can be used to distinguish between genuine entanglement and biseparable entanglement of pure states and can even specify the exact nature of biseparable entanglement$.\,\,$Finally we conjecture generalization of our protocol for detecting entanglement of multipartite ($m$$\geq$$4$) pure states. Apart from entanglement detection, the study of analyzing quantumness of network correlations may be contributory in study of various information processing tasks such as distribution of quantum key (QKD) \cite{key1,Mayer,Acin,key2}, generation of private randomness \cite{Colbeck,Pironio}, Bayesian game theoretic applications \cite{game}, etc.\\
Rest of our work is organized as follows. We start with discussing the motivation behind our work in sec.\ref{mot} followed by some basic preliminaries in sec.\ref{pre}. In sec.\ref{bip} first we analyze nature of quantum correlations generated in $n$-local linear network \cite{km1} using $n$ number of bipartite quantum states followed by proposal of the scheme for bipartite entanglement detection. In sec.\ref{trip}, first we derive the set of Bell inequalities for the non-linear trilocal network scenario, study violation of corresponding inequalities by pure tripartite quantum states  and then design tripartite entanglement detection scheme for some pure tripartite states. In sec.\ref{general}, we generalize the non-linear trilocal network to non-linear $n$-local network scenario when each of $n$ independent sources generates an $m$-partite ($m$$\geq$$4$) state. Finally we end with some concluding remarks in sec.\ref{conc}.\\
\section{Motivation}\label{mot}
Nonlocal behavior (Bell nonlocality) of correlations acts as a signature of presence of entanglement distributed (by a common source) among the parties who perform local measurements on their respective particles forming the entangled state. In network scenario, specifically for bilocal network Gisin \emph{et al.}  proved that all bipartite entangled states violate the bilocal inequality (see sec.\ref{pre}) indicating nonbilocality of corresponding network correlations \cite{BRAN}. Their findings \cite{ng1} generates the idea of using a bilocal network to detect entanglement of the states distributed by the sources. This idea basically motivates our present work. We exploit nonclassical nature of quantum correlations generated in a network (involving $n$ independent sources) where all the parties do not have access to random input selection. Subsequently we use the observations for detecting entanglement of bipartite states involved in the network. We not only confine within the scope of bipartite entanglement but consider tripartite entanglement also.
\section{Preliminaries}\label{pre}

\subsection{Bilocal Scenario}
Bilocal network (Fig.1) was framed in \cite{BRA,BRAN}. It is a network of three parties, say, Alice ($A$), Bob ($B$) and Charlie ($C$) and two sources $\mathcal{\textbf{S}}_1$ and $\mathcal{\textbf{S}}_2$ arranged linearly. Sources $\textbf{S}_1$ and $\textbf{S}_2$ are independent to each other ($\textit{bilocal assumption}$). Each of $\mathcal{\textbf{S}}_1$ and $\mathcal{\textbf{S}}_2$ sends a physical system characterized by variables $\lambda_1$ and $\lambda_2$ respectively. Intermediate party Bob gets two particles (one from each source). In $P^{14}$ scenario \cite{BRA,BRAN}, each of Alice and Charlie performs any one of two binary output measurements on their respective subsystems: $x,\,z\in\{0,1\}$ denote respective input sets for  Alice and Charlie whereas their outputs are labeled as $a,\,c\in\{0,1\}.$ Bob performs a single (fixed) measurement ($y$) having $4$ outcomes: $b$$=$$\overrightarrow{\mathbf{b}}$$=$$b_0b_1=00,01,10,11$ on the joint state of the two subsystems received from $\textbf{S}_1$ and $\textbf{S}_2$.\\
Correlations generated in the network are local if these can be decomposed as:
$\small{P_{14} (a, b, c|x, y, z)}=\iint \small{d\lambda_{1} d\lambda_{2} \rho (\lambda_1,\lambda_2)V}$
\begin{equation}\label{p11}
\small{\textmd{with}\,\,V}=\small{P_{14}(a|x, \lambda_1)P_{14}(b|y, \lambda_1, \lambda_2)P_{14}(c|z, \lambda_2)}
\end{equation}
Tripartite correlations $P_{14}(a, b, c|x, y, z)$ are bilocal if these can be written in above form (Eq.(\ref{p11}))
along with the constraint (referred to as \textit{bilocal constraint}):
\begin{equation}\label{p2}
    \rho(\lambda_1,\lambda_2)=\rho_1(\lambda_1)\rho_2(\lambda_2)
\end{equation}
imposed on the probability distributions of the hidden variables $\lambda_1, \lambda_2$. A linear extension of this model involving $n$ independent sources and $n$$+$$1$ parties was made in \cite{km1}.\\
Nonbilocality of tripartite correlations is guaranteed if these violate the inequality:$\sqrt{|I|} $$+$$ \sqrt{|J|}\leq1$ (for details, see \cite{BRAN}). Quantum violation of the bilocal inequality was pointed out in \cite{ng1}. Based on their findings it can be said that any bipartite two qubit entangled state violates the bilocal inequality.
\begin{figure}[htb]
\includegraphics[width=3in]{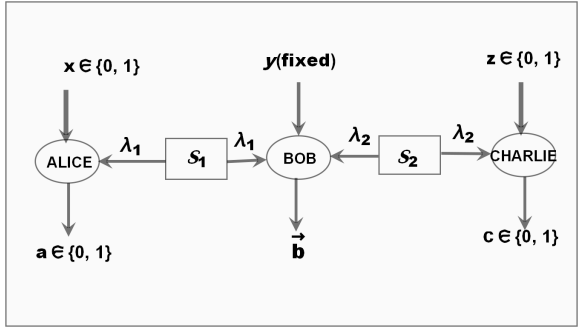}\\
\caption{\emph{Schematic diagram of bilocal network \cite{BRAN,BRA}. In $P_{14}$ scenario $y$ denotes fixed measurement of Bob together with $\vec{b}$ referring to $4$ outputs $ b_0b_1=00,\,01,\,10,\,11.$ }}
\end{figure}
Linear extension of bilocal network, referred to as $n$-\textit{local linear network} was given in \cite{km1} where the number of independent sources is $n.$ $n$-local quantum linear network is considered for our purpose (see Fig.2).
$\,$\\
\subsection{Complete Bell basis and GHZ basis Measurement}
Both of these measurements are instances of quantum entangled (joint) measurements. The operator of complete Bell basis measurement \cite{ng1}, often referred to as 'Bell state measurement' (BSM), is represented in terms of its four eigen vectors (Bell states):
\begin{equation}\label{bell1}
    |\phi^{\pm}\rangle=\frac{|00\rangle\pm|11\rangle}{\sqrt{2}}
\end{equation}
\begin{equation}\label{bell2}
    |\psi^{\pm}\rangle=\frac{|01\rangle\pm|10\rangle}{\sqrt{2}}
\end{equation}
Analogous to the bipartite entangled measurement of BSM, the operator corresponding to tripartite entangled measurement of complete GHZ basis measurement (GSM) is given in terms of the GHZ basis \cite{star}:
\begin{equation}\label{ap1}
|\phi_{mnk}\rangle_{GHZ}=\frac{1}{\sqrt{2}}\sum_{r=0}^1(-1)^{m*r}|r\rangle|r\small{\oplus} n \rangle |r\small{\oplus} k\rangle,\,m,n,k\in\{0,1\}.
\end{equation}
\section{Quantum Violation of Linear $n$-Local Inequality}\label{bip}
Here we consider an $n$-local linear network \cite{km1} involving quantum states (see Fig.2). Let each of $n$ independent sources generates a two qubit state: source $\mathcal{\textbf{S}}_i$ generating state $\varrho_i (i$$=$$1,2,...,n).$ Two qubits of state $\varrho_i$ are sent to parties $\mathcal{P}_{i}$ and $\mathcal{P}_{i+1} (i$$=$$1,2,...,n).$ The overall joint quantum system involved in the network is $\otimes_{i=1}^n\varrho_i.$ After receiving qubits, each of the extreme two parties $\mathcal{P}_1$ and $\mathcal{P}_{n+1}$ performs projective measurements in any of two arbitrary directions locally on their respective particles:$\mathcal{P}_1$ chooses any one of directions $\vec{\alpha_0}$ and $\vec{\alpha_1}$ (say) whereas for $\mathcal{P}_{n+1}$ let the directions be along any one $\vec{\beta_0}$ and $\vec{\beta_1}.$ Each of remaining $n-1$ intermediate parties $\mathcal{P}_i(i$$=$$2,...,n-1)$ performs a complete Bell-basis measurement (fixed setting) on the joint state of their respective two particles received from adjoining sources $\mathcal{\textbf{S}}_i$ and $\mathcal{\textbf{S}}_{i+1}$ (see Fig.2). $n$$+$$1$-partite correlations generated in the network are then used to test the $n$-local inequality \cite{km1}:
 \begin{equation}\label{A3}
 \sqrt{|I_{14}|} + \sqrt{|J_{14}|}\leq1.
\end{equation}
Terms appearing in above equation are detailed in Table.I.
 \begin{widetext}
\begin{center}
\begin{table}[htp]
\begin{center}
\begin{tabular}{|c|c|c|}
\hline
$I_{14}$ and $J_{14}$&Correlators&Measurement and \\
&&outputs of $\mathcal{P}_{i}(i=2,...,n)$\\
\hline
$I_{14}=\frac{1}{4}\sum \limits_{x_1,x_{n+1}=0,1}\langle E_1 E_2^0...E_n^0E_{n+1}\rangle$ &$\langle E_1 E_2^y...E_n^yE_{n+1}\rangle=\sum\limits_U (-1)^{a_1+a_{2y}+...+a_{ny}+a_{n+1}}V_{14}$&$E_i^ y$:\small{observable of} $\mathcal{P}_i$\\
&$U=\{a_1, a_{20},a_{21},...,a_{n0},a_{n1},a_{n+1}\}$&\small{corresponding to a single input} \\
$J_{14}=\frac{1}{4}\sum \limits_{x_1,x_{n+1}=0,1}(-1)^{x_1+x_{n+1}}\langle E_1 E_2^1...E_n^1E_{n+1}\rangle$&$V_{14}=P(a_1, \overrightarrow{a_2},...,\overrightarrow{a_n},a_{n+1}|x_1,x_{n+1})$& having $4$ \small{outputs:}\\
&& $\overrightarrow{a_i}=a_{i0}a_{i1}=00,\,01,\,10,\,11.$\\
\hline
\end{tabular}\\
\caption{Details of the terms appearing in Eq.(\ref{A3}). $E_1$ and $E_{n+1}$ denote respective observables corresponding to binary inputs $x_1$ and $x_{n+1}$ of parties $\mathcal{P}_1$ and $\mathcal{P}_{n+1}$. $a_1,a_{n+1}\in\{0,1\}$ stand for corresponding outputs. }
\end{center}
  	\label{table1}
  \end{table}
  \end{center}
  \begin{figure}[htb]
\includegraphics[width=3.6in]{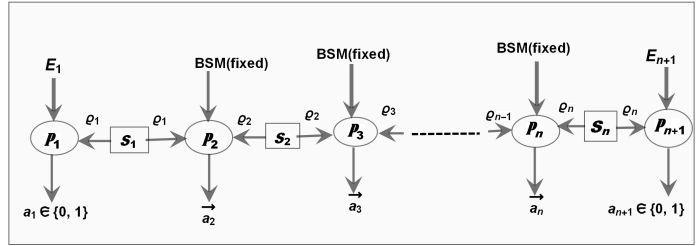}\\
\caption{\emph{Schematic diagram of quantum $n$-local linear network. $E_1$ and $E_{n+1}$ stand for observables corresponding to binary inputs $x_1$ and $x_{n+1}$ of $\mathcal{P}_1$ and $\mathcal{P}_{n+1}$ respectively. Here each of $\mathcal{P}_i(i$$=$$2,...,n)$ performs single measurement denoted by 'BSM' which stands for complete Bell basis measurement with $\vec{a_i}$ referring to $4$ outputs $ a_{i0}a_{i1}=00,\,01,\,10,\,11.$ . }}
\end{figure}
\end{widetext}
 Clearly, excepting the extreme two parties $A_1$ and $A_{n+1},$ none of the remaining $n$$-$$1$ parties have access to random choice of measurements. Under such circumstances, we consider two separate cases.\\
\textit{Network involving pure states:} Let each of the $n$ sources generates a pure two qubit state. To be precise, say $\mathcal{\textbf{S}}_i$ emits:
\begin{equation}\label{st1}
    \varrho_i=\gamma_{0i}|00\rangle+\gamma_{1i}|11\rangle
\end{equation}
where $\gamma_{0i}$ and $\gamma_{1i}(i$$=$$1,...,n)$ are positive real Schmidt coefficients \cite{preskil} satisfying normalization condition $\gamma_{oi}^2$$+$$\gamma_{1i}^2$$=$$1.$ $\varrho_i$ is entangled for any non zero value of both $\gamma_{0i}$ and $\gamma_{i1}\,(i$$=$$1,2,...,n),$ i.e., $\gamma_{0i}\gamma_{1i}$$>$$0.\,$Maximizing over all possible projective measurement directions of extreme two parties $A_1$ and $A_{n+1}$, the upper bound of violation ($\mathcal{B}_{14}$,say) of Eq.(\ref{A3}) turns out to be:
\begin{equation}\label{st2}
    \mathcal{B}_{14}=\mathcal{B}_{14}^{(pure)}=\sqrt{1+2^n\Pi_{i=1}^n\gamma_{oi}\gamma_{1i}}.
\end{equation}
$\mathcal{B}_{14}$$>$$1$ implies that all the pure states involved in the network are entangled. Hence, up to the existing sufficient criterion given by Eq.(\ref{A3}) for detecting nonbilocality, nonbilocal correlations are generated in a network only if all pure states involved in the network are entangled.\\
\textit{Network involving mixed bipartite states:} Let us now consider the case when each of $\mathcal{\textbf{S}}_j$ emits a mixed bipartite state (density matrix formalism):
\begin{equation}\label{st1}
    \varrho_j=\frac{1}{2^2}\sum_{i_1,i_2=0}^{3}t_{i_1i_2}^{(j)}\sigma^1_{i_1}\bigotimes\sigma^2_{i_2}
\end{equation}
where $\sigma^q_0,$ stands for the identity operator of the Hilbert space which is associated with qubit $q$ and $\sigma^q_{i_q},$ denote the Pauli operators along three mutually perpendicular directions, $i_q$$=$$1,2,3$. $t_{i_1i_2}^{(j)}(i,j$$=$$1,2,3)$ denote the elements of the correlation tensor $T^{(j)}$ (say) of the bipartite state $\varrho_j.$ Polar value decomposition of correlation tensor ($T^{(j)}$) for each of $\varrho_j$ generates the matrix $U^{(j)}M^{(j)}$$=$$T^{(j)}$ where $U^{(j)}$ denotes a unitary matrix and $M^{(j)}$$=$$\sqrt{(T^{(j)})^{\dagger}T^{(j)}}$ having eigen values $\lambda_1^{(j)}$$\geq$$\lambda_2^{(j)}$$\geq$$\lambda_3^{(j)}$. The polar decomposition of $\varrho^{(j)}$ and $\varrho^{(j+1)}$ characterize the fixed measurement (BSM) of $A_{j}\,(j$$=$$2,...,n)$ who performs suitable local unitaries over subsystems received from sources $\mathcal{\textbf{S}}_j$ and $\mathcal{\textbf{S}}_{j+1}$ (for detailed discussion on the methodology used here,see \cite{ng1}). The upper bound of violation ($\mathcal{B}_{14}$) now turns out to be:
\begin{equation}\label{st2i}
    \mathcal{B}_{14}=\mathcal{B}_{14}^{(mixed)}=\sqrt{\Pi_{j=1}^{n}\lambda_1^{(j)}+\Pi_{j=1}^{n}\lambda_2^{(j)}}.
\end{equation}
Now, let none of $\varrho_j\,(j=1,...,n)$ violates standard Bell-CHSH inequality, i.e., by Horodecki criterion \cite{HOR}:
\begin{equation}\label{crit1}
   \mathcal{B}_{CHSH}^{(j)}= \sqrt{(\lambda_1^{(j)})^2+(\lambda_2^{(j)})^2}\leq 1
\end{equation}
where $\mathcal{B}_{CHSH}^{(j)}$ denotes the upper bound of violation of Bell-CHSH inequality by $\varrho_j.$ This in turn indicates that for each of $\varrho_j(j$$=$$1,...,n),$ $\lambda_i^{(j)}(i$$=$$1,2,3)$$<$$1.$ Under such circumstances, Eq.(\ref{st2i}) gives:
\begin{eqnarray}
\mathcal{B}_{14}&<& \textmd{max}_{k\neq l}\sqrt{\lambda_1^{(k)}\lambda_1^{(l)}+\lambda_2^{(k)}\lambda_2^{(l)}},\,k,l=1,...,n  \nonumber\\
                &\leq&\sqrt{(\lambda_1^{(k)})^2+(\lambda_2^{(k)})^2}\,\sqrt{(\lambda_1^{(l)})^2+(\lambda_2^{(l)})^2} \nonumber\\
                &=& \mathcal{B}_{CHSH}^{(k)}  \mathcal{B}_{CHSH}^{(l)}\nonumber\\
                &\leq&1.
\end{eqnarray}
Hence, $\mathcal{B}_{14}$$>$$1$ implies that at least one of the states $\varrho_j$ generated by $\mathcal{\textbf{S}}_j$ is Bell-CHSH nonlocal.

\subsection{Bipartite Entanglement Detection}
Let there be $n$ unknown bipartite quantum states $\Phi_i$ generated by $n$ distinct sources $\mathcal{\textbf{S}}_i(i$$=$$1,...,n).$ All these $n$ sources being spatially separated, they are independent of each other. In order to detect whether at least one of $\Phi_i$ is entangled or not, let the sources be arranged linearly and the states be distributed among $n$$+$$1$ parties $\mathcal{P}_i(i$$=$$1,...,n$$+$$1)$  so as to form a $n$-local network (Fig.2). Let each of $\mathcal{P}_1$ and $\mathcal{P}_{n+1}$ performs projective measurements in anyone of two arbitrary directions whereas intermediate $n-1$ parties (receiving two particles each) perform complete Bell basis measurement (BSM). Practical implementation of this protocol, where only some of the parties ($\mathcal{P}_1,\mathcal{P}_{n+1}$) have to choose randomly from a set of two measurements, is easier compared to any protocol where none of the parties involved performs fixed measurement. $n$$+$$1$-partite correlations generated therein are used to test the $n$-local inequality (Eq.(\ref{A3})). Observation of violation of the inequality guarantees that at least one of $\Phi_i$ is entangled$.\,\,$Utility of the violation of Eq.(\ref{A3}) is already justified in the previous subsection. Clearly, this protocol detects entanglement of all the states involved in a device dependent manner (as each of the intermediate parties have to perform BSM  thereby making the scheme depending on inner working of the device) in case all $\Phi_i(i$$=$$1,...,n)$ are identical copies of an unknown quantum state. \\
Having used a $n$-local linear network for the purpose of bipartite entanglement detection, we now proceed to do the same for some families of pure tripartite entangled states. For that, we first analyze trilocal non-linear network scenario.
\section{Trilocal Non-Linear Network Scenario}\label{trip}
The scenario is based on a five party ($\mathcal{P}_1^E,\mathcal{P}_2^E,\mathcal{P}_3^E,\mathcal{P}_1^I,\mathcal{P}_2^I$) network involving three independent sources $\mathcal{\textbf{S}}_1$, $\mathcal{\textbf{S}}_2$ and $\mathcal{\textbf{S}}_3$ (see Fig.3). Source $\mathcal{\textbf{S}}_i$ is characterized by hidden variable $\eta_i(i$$=$$1,2,3).$ Source independence implies existence of independent probability distributions:
 \begin{equation}\label{n6}
    \Lambda(\eta_1,\eta_2,\eta_3)=\Lambda_1(\eta_1)\Lambda_2(\eta_2)\Lambda_3(\eta_3)
\end{equation}
where $\int d\eta_i \Lambda_i(\eta_i)$$=$$1\forall i$. Source $\mathcal{\textbf{S}}_i$ sends particles to parties $\mathcal{P}_{1}^I,\,\mathcal{P}_{2}^I,\,\mathcal{P}_i^E\,(i$$=$$1,2,3).$ Parties $\mathcal{P}_1^I$ and $\mathcal{P}_2^I$ receiving three particles (one from each source) are referred to as \textit{intermediate} parties and remaining three parties $\mathcal{P}_1^E,\mathcal{P}_2^E,\mathcal{P}_3^E,$ each receiving a single particle, are referred to as \textit{extreme} parties. Let $x_1,\,x_2,\,x_3($$\in$$\{0,1\})$ stand for binary inputs of parties $\mathcal{P}_1^E,\,\mathcal{P}_2^E,\,\mathcal{P}_3^E$ respectively whereas $a_1,\,a_2,\,a_3($$\in$$\{0,1\})$ correspond to the respective outputs. Each of $\mathcal{P}_1^I$ and $\mathcal{P}_2^I$ has access to single input giving eight outputs: $\vec{b_1}$$=$$(b_{10},b_{11},b_{12})$ and  $\vec{b_2}$$=$$(b_{20},b_{21},b_{22})(b_{ij}$$\in$$\{0,1\}\forall\,i$$=$$1,2\,\textmd{and}\,j$$\in$$\{0,1,2\})$ denote the outputs of parties $\mathcal{P}_1^I$ and $ \mathcal{P}_2^I$ respectively. Parties are not allowed to communicate between themselves. Correlations generated in this network scenario are \textit{trilocal} if they can be factorized as follows:
\begin{widetext}
$P_{18}(a_1,\vec{b_1},\vec{b_2},a_2,a_3|x_1,x_2,x_3)$$ =$$ \iiint d\eta_{1} d\eta_{2}d\eta_{3} {\Lambda(\eta_1,\eta_2,\eta_3)}W$ where
\begin{equation}\label{n7}
W=P_{18}(a_1|x_1, \eta_1)P_{18}(\vec{b_1}| \eta_1, \eta_2,\eta_3)P_{18}(\vec{b_2}| \eta_1, \eta_2,\eta_3)P_{18}(a_2|x_2, \eta_2)P_{18}(a_3|x_3, \eta_3)
\end{equation}
\end{widetext}
along with the restriction imposed by Eq.(\ref{n6}). Under source independence restriction (Eq.(\ref{n6})), correlations which cannot be decomposed as above (Eq.(\ref{n7})) are said to be \textit{nontrilocal} in nature. It may be noted that the network scenario introduced here is in some extent similar to that of the scenario discussed in \cite{km2} where each of the parties involved has the freedom to choose from a set of two measurements. So the scenario in present discussion and that introduced in \cite{km2} differ on the basis of whether the intermediate parties perform a single measurement or not. Correspondingly the correlations characterizing the measurement scenarios and the inequalities involved therein are different from those discussed in \cite{km2}. We now derive a set of sufficient criteria in the form of non-linear Bell-type inequalities sufficient to detect nontrilocal correlations.\\
\textit{Theorem.1}: For any trilocal five partite correlation, each of the following inequalities necessarily holds:
\begin{equation}\label{n8t}
    \sqrt[3]{|I_{m_1,m_2,0}^{(18)}|}+\sqrt[3]{|I_{n_1,n_2,1}^{(18)}|}\leq1\,\,\forall\,m_1,\,m_2,\,n_1,\,n_2\,\in\{0,1\}
\end{equation}
For details of the correlators used in Eq.(\ref{n8t}), see Table.II.\\
\textit{Proof:} For proof, see Appendix.A.\\
The set of $16$ inequalities given by Eq.(\ref{n8t}) being only necessary criteria of trilocality, there may exist nontrilocal correlations satisfying all of them. However, violation of at least one of these inequalities guarantees nontrilocality of the correlations. Violation of Eq.(\ref{n8t}) for at least one possible $(m_1,m_2,n_1,n_2)$ is thus sufficient for detecting nontrilocality of corresponding correlations.
 \begin{widetext}
\begin{center}
\begin{table}[htp]
\begin{center}
\begin{tabular}{|c|}
\hline
Correlators \\
\hline
$I_{m_1(n_1),m_2(n_2),i}^{(18)}=\frac{1}{8}\sum \limits_{x_1,x_2,x_3=0,1}(-1)^{i*(x_1+x_2+x_3)}\langle A_{1,x_1} A_{2}^{m_1(n_1)} A_{3}^{m_2(n_2)}  A_{4,x_2} A_{5,x_3}\rangle,\,i,\,m_1,\,m_2,\,n_1,\,n_2\,\in\{0,1\}$\\
$\,$\\
\hline
 $\langle A_{1,x_1} B_{1}^{m_1(n_1)} B_{2}^{m_2(n_2)}  A_{2,x_2} A_{3,x_3}\rangle$$=$$\sum\limits_{\mathcal{C}}(-1)^{h}P_{18}(a_1,\vec{b_1},\vec{b_2},a_2,a_3|x_1,x_2,x_3)$ \\
  where $\mathcal{C}$$=$$\{a_1,a_2,a_3,b_{10},b_{11},b_{12}, b_{20},b_{21},b_{22}\}$ and $h=a_1$$+$$a_2$$+$$a_3$$+$$s_{m_1(n_1)}(b_{10},b_{11},b_{12})$$+$$s_{m_2(n_2)}(b_{20},b_{21},b_{22})$\\\
  $\,$\\
 \hline
 with functions $s_i(x,y,z)$ being defined as $s_0(x,y,z)$$=$$x$$+$$y$$+$$z$$+$$1$ and $s_1(x,y,z)$$=$$x*y$$+$$y*z$$+$$x*z$\\
 $\,$\\
  \hline
\end{tabular}\\
\caption{Detailing of the terms used in Eq.(\ref{n8t}). $A_i$ denote the observable for input $x_i$ of $\mathcal{P}_i^E(i$$=$$1,2,3)$ whereas $B_1,B_2$ denote observable corresponding to single input of $\mathcal{P}_1^I$ and $\mathcal{P}_2^I$ respectively.}
\end{center}
  	\label{table2}
  \end{table}
  \end{center}
  \begin{figure}[htb]
\includegraphics[width=4in]{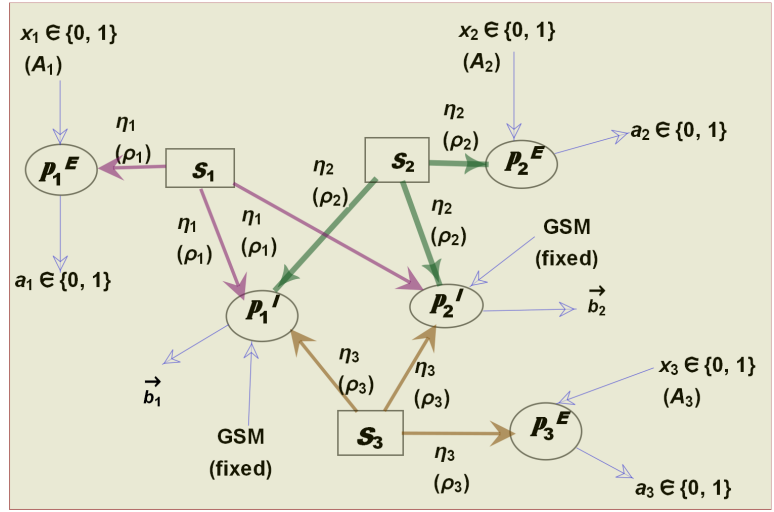}\\
\caption{\emph{Trilocal non-linear network. Source $\mathcal{\textbf{S}}_i$ is characterized by hidden variable $\eta_i\,(i$$=$$1,2,3).$ In case of quantum network $\mathcal{\textbf{S}}_i$ generates tripartite quantum state $\rho_i.$ $A_i$ denotes observables corresponding to binary inputs $x_i$ of $\mathcal{P}_i\,($$=$$1,2,3)$ respectively. Here each of $\mathcal{P}_1^I,\,\mathcal{P}_2^I$ performs single measurement (GHZ basis measurement) denoted by 'GSM' with three dimensional vector $\vec{b_i}$ now referring to $8$ outputs  $b_{i0}b_{i1}b_{i2}{=000,\,001,\,010,\,100,\,101,\,110,\,011,\,111}.$  }}
\end{figure}
\begin{center}
\begin{table}[htp]
\begin{center}
\begin{tabular}{|c|c|}
\hline
State($\rho_i$ generated  &State Parameters giving  \\
by $\mathcal{\textbf{S}}_i$)&violation \\
\hline
$ |\varphi_{\small{GHZ}}^{(i)}\rangle$(Eq.(\ref{S15}))&$\beta_1$$=$$0.72$, $\beta_2$$=$$0.75$, $\beta_3$$=$$0.7$\\
$\,$&$\,$\\
 \hline
$|\varphi_{W}^{(i)}\rangle$(Eq.(\ref{wst}))&$\omega_{1i}$$=$$0.558327$, $\omega_{2i}$$=$$1.5708,\,\forall\,i\in\{1,2,3\}$\\
$\,$&$\,$\\
\hline
$|\varphi^i_{(12|3)}\rangle$(Eq.(\ref{bisep1}))&$c_{0i}$$=$$0.592368,\,v_{0i}$$=$$1,\,\forall\, i\in\{1,2,3\}$\\
\hline
$|\varphi^i_{(13|2)}\rangle$(Eq.(\ref{bisep2}))&$c_{0i}$$=$$1.5708$$-$$\imath\,0.15776,\,v_{0i}$$=$$1,\,\forall\, i\in\{1,2,3\}$\\
\hline
$|\varphi^i_{(23|1)}\rangle$(Eq.(\ref{bisep3}))&\small{no violation is obtained.upper bound($\mathcal{B}_{18},$say)}\\
$\,$& \small{of trilocal inequalities(Eq.(\ref{n8t}))} \\
 $\,$&\small{for identical copies,}$\mathcal{B}_{18}$$=$$\textmd{Max}[2^{\frac{2}{3}}|c_{01}c_{11}|,(c_{01}^4$$+$$4c_{01}^3c_{11}^3$$+$$c_{11}^4)^{\frac{1}{3}}]¸$\\
$\,$& where $c_{k1}$$=$$c_{k2}$$=$$c_{k3},$ $k$$=$$0,1.$\\
\hline
\end{tabular}\\
\caption{Exploring some specific instances of nontrilocal nature of correlations observed when some tripartite pure quantum states are used in the non-linear trilocal network under the assumption that each of $\mathcal{P}_1^E,$ $\mathcal{P}_2^E$ and $\mathcal{P}_3^E$ receives $1^{st}$ qubit of $\rho_1,$ $\rho_2$ and $\rho_3$ respectively whereas remaining two qubits of each $\rho_i$ are received by the intermediate parties. No violation of trilocal inequalities (at least one) is however obtained when tripartite state has entangled $2^{nd}$ and $3^{rd}$ qubits (Eq.(\ref{bisep3})) as $\mathcal{B}_{18}$$\leq$$1$.}
\end{center}
  	\label{table3}
  \end{table}
  \end{center}
\end{widetext}
\subsection{Quantum Violation}\label{qu}
Consider a network involving three independent sources $\mathcal{\textbf{S}}_1$, $\mathcal{\textbf{S}}_2$ and $\mathcal{\textbf{S}}_3$ each generating a three qubit state $\rho^{(i)}$ (see Fig.3). The overall quantum state involved in the network becomes:
\begin{equation}\label{n13}
\rho_{12345}=\rho^{(1)}\small{\otimes}\rho^{(2)}\small{\otimes}\rho^{(3)}.
\end{equation}
After the qubits are distributed from the sources, no communication takes place between the parties who now perform measurements on their respective subsystems. Each of $\mathcal{P}_1^I$ and $\mathcal{P}_2^I$ performs complete GHZ basis measurement (GSM) on the joint state of the three qubits that each of them receives from the three sources. Each of $\mathcal{P}_1^E,$ $\mathcal{P}_2^E$ and $\mathcal{P}_3^E$ performs projective measurements on their single qubit in any of two arbitrary directions: $\mathcal{P}_i^I(i$$=$$1,2,3)$ measures in anyone of $\vec{\gamma}_{i0}$ and $\vec{\gamma}_{i1}$ directions . \\
$\textmd{Interestingly, if each of the sources}\,\mathcal{\textbf{S}}_i\,\textmd{generates arbitrary}$\\
$\textmd{tripartite product state}$:
\begin{equation}\label{product}
    \rho_i=\otimes_{j=1}^3(v_{0ij}|0\rangle+v_{1ij}|1\rangle)(|v_{0ij}|^2+|v_{1ij}|^2=1),
\end{equation}
none of the inequalities given by Eq.(\ref{n8t}) is violated. We now proceed to discuss some possible cases of quantum violation of inequalities given by Eq.(\ref{n8t}). For our purpose we consider tripartite pure states. \\
Let each of the sources generates an arbitrary biseparable (in $12$ by $3$ cut) entangled state:
\begin{equation}\label{bisep1}
  |\varphi^i_{(12|3)}\rangle=(c_{0i}|00\rangle_{12}+c_{1i}|11\rangle_{12}) \otimes(v_{0i}|0\rangle_{3}+v_{1i}|1\rangle_{3})
\end{equation}
with$\,v_{0i}^2$$+$$v_{1i}^2$$=$$1$ and $c_{0i}^2$$+$$c_{1i}^2$$=$$1(v_{ij},\, c_{ij}\,\textmd{are the}$
Schmidt coefficients) \cite{preskil}. Now, compatible with the arrangement of the sources and parties in this network, let $1^{st}$ qubit of each $\rho_i$$=$$|\varphi^i_{(12|3)}\rangle\langle\varphi^i_{(12|3)}|(i$$=$$1,2,3)$ is sent to the extreme parties: $\mathcal{P}_1^E$, $\mathcal{P}_2^E$ and $\mathcal{P}_3^E$ receiving $1^{st}$ qubit of $\rho_1$, $\rho_2$ and $\rho_3$ respectively whereas $2^{nd}$ and $3^{rd}$ qubits of each $\rho_i$ are sent to the intermediate parties: $\mathcal{P}_1^I$ receives $2^{nd}$ qubit of $\rho_1,\,\rho_2,\,\rho_3$ and $\mathcal{P}_{3}^I$ receives $3^{rd}$ qubit of these states. Violation of Eq.(\ref{n8t}) is observed for some members of this family (Eq.(\ref{bisep1})). Violation is also observed if each of $\mathcal{\textbf{S}}_i$ generates some states having biseparable entanglement in $13$ by $2$ cut:
\begin{equation}\label{bisep2}
  |\varphi^i_{(13|2)}\rangle=(c_{0i}|00\rangle_{13}+c_{1i}|11\rangle_{13}) \otimes(v_{0i}|0\rangle_{2}+v_{1i}|1\rangle_{2})
\end{equation}
However, violation is impossible if $\mathcal{\textbf{S}}_i$ generates any member from the family of biseparable entangled states having entanglement among its $2^{nd}$ and $3^{rd}$ qubits:
\begin{equation}\label{bisep3}
   |\varphi^i_{(23|1)}\rangle=(c_{0i}|00\rangle_{23}+c_{1i}|11\rangle_{23}) \otimes(v_{0i}|0\rangle_{1}+v_{1i}|1\rangle_{1})
\end{equation}
At this junction it should be noted that violation of Eq.\ref{n8t}) depends on the order of distribution of qubits of each $\rho_i\,(i=1,2,3)$ among the parties. Compatible with the network scenario (Fig.3), when $1^{st}$ qubit of each $\rho_i\,(i$$=$$1,2,3)$ is sent to the extreme parties and remaining two qubits of each $\rho_i$ are received by the intermediate parties (as discussed), violation is observed in networks involving biseparable entanglement in $13$ by $2$ (Eq.(\ref{bisep2})) or $12$ by $3$ (Eq.(\ref{bisep1})) cuts only. But violation is not observed if $\rho_i$ have biseparable entanglement in $23$ by $1$ cut(Eq.(\ref{bisep3}). But networks involving biseparable entanglement in $23$ by $1$ (Eq.(\ref{bisep3})) cut also gives violation if $2^{nd}$ qubit of each $\rho_i\,(i$$=$$1,2,3)$ is sent to the extreme parties and remaining two qubits of each $\rho_i$ are received by the intermediate parties. However, violation of any one of the trilocal inequalities given by Eq.(\ref{n8t}) is not always arrangement(of qubits) specific. We consider genuine entanglement in support of our claim.\\
Let each of $\mathcal{\textbf{S}}_i$ in the non-linear trilocal network now generates a Generalized GHZ (GGHZ) state (\cite{ACN}), $\rho_i$$=$$|\varphi_{\small{GHZ}}^{(i)}\rangle\langle \varphi_{\small{GHZ}}^{(i)}|$ where,
\begin{equation}\label{S15}
 |\varphi_{\small{GHZ}}^{(i)}\rangle\,=\,\cos(\beta_i)|000\rangle+\sin(\beta_i)|111\rangle,\,\beta_i\in[0,\frac{\pi}{4}].
\end{equation}
Contrary to biseparable entanglement, nontrilocal correlations are obtained in the network for some states from the GGHZ family (Eq.(\ref{S15})) irrespective of distribution of qubits of each of the states ($\rho_i$). Analogous observation is obtained when $W$ states \cite{ACN5} are involved in the network:\\
$ |\varphi_{W}^{(i)}\rangle$$=$$  \cos\omega_{2i}\sin\omega_{1i}|001\rangle$$+$$\sin\omega_{2i}\sin\omega_{1i}|010\rangle$$+$
\begin{equation}\label{wst}
     \cos\omega_{1i}|100\rangle,\,\omega_{1i},\,\omega_{2i}\in[0,\frac{\pi}{4}].
\end{equation}
Now, if both biseparable and genuine entanglement of W state (Eq.(\ref{wst})) are used in the network, violation again depends on arrangement of qubits. Here it should be pointed out that if one of the three tripartite pure states generated by the sources is a product state then violation of trilocal inequalities cannot be observed even if the remaining two states are entangled.\\
Based on the above analysis of quantum violation and the fact that such violation is sufficient to detect nontrilocal nature of network correlations, we now design a scheme to detect both biseparable and genuine entanglement of tripartite pure states. But it may be pointed out that this scheme may fail to detect presence of entanglement in some cases as violation is not possible for all tripartite pure entangled states. \\
\subsection{Tripartite Pure Entanglement Detection}
Consider a non-linear trilocal network. Let three unknown pure tripartite states $\kappa_1,$ $\kappa_2$ and $\kappa_3$ are generated by $\mathcal{\textbf{S}}_1,$ $\mathcal{\textbf{S}}_2$ and $\mathcal{\textbf{S}}_3$ in the network. Distribution of qubits among the parties plays a significant role in violation of trilocal inequalities. Consequently for designing a scheme of entanglement detection, we consider all the possible arrangement of qubits. The protocol breaks up into $27$ phases: $t_{i,j,k}(i,j,k\in\{1,2,3\}).$ In phase $t_{i,j,k},$  for every possible value of $i,j,k\in\{1,2,3\},$ $i^{th},j^{th},k^{th}$ qubit of $\kappa_1,\kappa_2,\kappa_3$ respectively are sent to extreme parties. Hence $\mathcal{P}_1^{E},$ $\mathcal{P}_2^{E}$ and $\mathcal{P}_3^{E}$ receives $i^{th},j^{th},k^{th}$ qubit of $\kappa_1,\kappa_2,\kappa_3$ respectively (for more details see Table.VI in Appendix.B)$.\,$Remaining qubits of each of the unknown states are distributed among the intermediate parties in any pattern compatible with the non-linear trilocal network scenario (Fig.3). One may note that ordering of the phases is not essential. After receiving the particles, in each of these phases, the parties perform measurements on their respective subsystems. Correlated statistics are then used to test the trilocal inequalities (Eq.(\ref{n8t})). If violation of at least one of the inequalities is observed in at least one phase then each of $\kappa_1,\,\kappa_2$ and $\kappa_3$ is a tripartite entangled state whereas violation in all the phases ensures genuine entanglement of all the three unknown states. In the protocol, either violation occurs in no phase or in specific number of phases (see Table.IV) \\
\begin{table}[htp]
\begin{center}
\begin{tabular}{|c|c|}
\hline
Total number&Implication\\
 of phases&\\
\hline
$0$&No definite conclusion. \\
\hline
$8$& All are biseparable.\\
\hline
$12$& Any two of the unknown \\
&states are biseparable and \\
&the remaining is genuinely entangled\\
& other than GGHZ (Eq.(\ref{S15})) or W (Eq.(\ref{wst})) classes.\\
\hline
$18$& Only one of three unknown\\
 &states is biseparable with\\
 & the other two being \\
 &genuinely entangled but does\\
 & not belong to GGHZ or W families.\\
\hline
$\geq$$19$ & each $\kappa_i$ has genuine\\
but $<$$27$&  entanglement but is neither a member \\
& of GGHZ family (Eq.(\ref{S15})) nor W state (Eq.(\ref{wst}))\\
\hline
$27$& Each of $\kappa_i$ is genuinely entangled.\\
\hline\end{tabular}\\
\caption{Total count of phases for which violation can be observed in the protocol is enlisted here. Implications are obvious from observations discussed in Section.VA. In case of no violation in any of the phases, the protocol fails to detect entanglement.  }
\end{center}
  	\label{table4}
  \end{table}
Interestingly, comparison of the possible nature of biseparable entanglement of $\kappa_1,\kappa_2,\kappa_3$ from Table.VI ensures nature of entanglement of each of the unknown states. To be more precise, at the end of the protocol, one can detect which of the three unknown states is genuinely entangled and which one is biseparable. Also the specific nature of biseparable entanglement can be detected.\\
As already discussed, total count of phases in which violation may be encountered is not arbitrary (see Table.IV). Leaving aside the implications in last two cases (corresponding to last two rows of Table.IV), let us consider the remaining cases individually:
\begin{itemize}
\item Let violation be obtained in $18$ phases. Then definitely two of three unknown states are genuinely entangled but is neither a GGHZ nor W state and the remaining one is a biseparable entangled state. For instance, violation in only first $18$ phases of the protocol ($t_{1,j,k},t_{2,j,k}, j,k$$\in$$\{1,2,3\}$) ensures that only $\kappa_1$ is a biseparable entangled state having entanglement in $12/3$ cut. This implication is obvious if one note that $12/3$ cut biseparable entanglement is the only possible nature of entanglement of $\kappa_1$ if violation is obtained in first $18$ phases (Table.VI). \\
\item Violation in only $12$ phases ensures that two of three unknown states are biseparable entangled and other one is genuinely entangled (other than GGHZ or W state). Consider a specific instance. Let violation be obtained in $t_{1,2,k},t_{3,2,k}, t_{1,3,k},t_{3,3,k}, \forall k$$\in$$\{1,2,3\}.$ Then $\kappa_1, \kappa_2$ are biseparable entangled states in $13/2$  and $23/1$ cut respectively and $\kappa_3$ is genuinely entangled.  \\
\item Violation in only $8$ phases ensures that all three unknown states are biseparable entangled. Nature of biseparable entanglement of each $\kappa_i$ is also detected. Consider the instance where violation is obtained in phases
$t_{1,2,k},t_{3,2,k}, t_{1,3,k},t_{3,3,k}, \forall k$$\in$$\{1,2\}.$ Then $\kappa_1,\kappa_2,\kappa_3$ are biseparable entangled states in $13/2,$ $23/1$ and $12/3$ cut respectively.\\
\end{itemize}
All these implications are direct consequences of the fact that violation of trilocal inequalities is not distribution (of qubits) specific in networks involving only genuine entanglement of GGHZ or W states whereas the same is crucial if at least one of the sources generates biseparable entanglement or genuine entanglement other than GGHZ (Eq.(\ref{S15}) and W (Eq.(\ref{wst}))classes.\\
\section{$n$-local Non Linear Network Scenario}\label{general}
Trilocal non linear network can be extended to a network involving $2n$$-$$1$ parties and $n$ independent sources, each generating an $n$ partite state. Each of $n$ number of parties $\mathcal{P}_i^E(i$$=$$1,2,...,n)$ (say) receives only one particle and are referred to as  \textit{extreme} parties whereas each of remaining $n-1$ parties $\mathcal{P}_i^I(i$$=$$1,2,...,n-1)$, referred to as \textit{intermediate} party, receives $n$ particles (each from one source). Let $x_i$$\in$$\{0,1\}$ and  $a_i$$\in$$\{0,1\}$ denote the binary input and output respectively of $\mathcal{A}_i(i$$=$$1,2,...,n).$ Each of $\mathcal{B}_i(i$$=$$1,2,...,n$$-$$1)$ performs a fixed measurement having $2^n$ outputs labeled as a $n$-dimensional vector $\vec{b_i}$$=$$(b_{i0},...,b_{in-1})$. After receiving qubits from the sources, parties do not communicate. $2n$$-$$1$ partite correlations are $n$-local if they can be decomposed as:
\begin{widetext}
$P_{12^n}(a_1,\vec{b_1},...,\vec{b}_{n-1},a_2,...,a_n|x_1,...,x_n) = \int...\int d\eta_{1}... d\eta_{n}{\Lambda(\eta_1,...,\eta_n)}W_n$ where
\begin{equation}\label{n7}
W_n=\Pi_{i=1}^nP_{12^n}(a_i| x_i,\eta_i)\Pi_{i=1}^{n-1}P_{12^n}(\vec{b_i}| \eta_1,..., \eta_n)
\end{equation}
together with the constraint:
\begin{equation}\label{nn}
    \Lambda(\eta_1,\eta_2,...,\eta_n)=\Pi_{i=1}^n\Lambda_i(\eta_i)
\end{equation}
\end{widetext}
where $\eta_i$ characterizes source $\mathcal{\textbf{S}}_i$ and $\int d\eta_i \Lambda_i(\eta_i)$$=$$1\,\forall\,i$$\in$$\{1,...,n\}.$
Correlations inexplicable in above form are non $n$-local. The $n$-local inequalities are given by the following theorem.\\
 \textit{Theorem.2}: Any $n$-local $2n$$-$$1$ partite correlation term necessarily satisfies:
\begin{equation}\label{n8tt}
    \sqrt[n]{|I_{f_1,...,f_{n-1},0}^{(12^n)}|}+\sqrt[n]{|I_{g_1,...,g_{n-1},1}^{(12^n)}|}\leq1
\end{equation}
where $ f_1,...,\,f_{n-1},\,g_1,..,\,g_{n-1}$$\in$$\{0,1\}.$\\
Correlators used in Eq.(\ref{n8tt}) are detailed in Table.V.\\
\textit{Proof:} Proof is based on the same technique as adopted for proving Theorem.1. As mentioned in Appendix.A, for proving Theorem.1 we need to relate the correlators (used in present scenario) with that introduced for designing another trilocal network scenario in \cite{km2}. Analogously Theorem.2 can be proved following the same line of argument (as that in Theorem.1). For that one should relate correlators (Table.V) introduced for the $n$-local non-linear scenario here with that of $n$-local network developed in \cite{km2}.
 \begin{widetext}
\begin{center}
\begin{table}[htp]
\begin{center}
\begin{tabular}{|c|}
\hline
Correlators related to $n$-local non-linear inequalities (Eq.(\ref{n8tt}))  \\
\hline
$I_{f_1(g_1)...,f_{n-1}(g_{n-1}),i}^{(12^n)}=\frac{1}{2^n}\sum \limits_{x_1,...,x_n=0,1}(-1)^{i*(x_1+...+x_n)}\langle A_{1,x_1} B_{1}^{f_1(g_1)} ...B_{n-1}^{f_{n-1}(g_{n-1})}...A_{n,x_n}\rangle,\,\textmd{with}\,i,\,f_1,...,\,f_{n-1},\,g_1,...,\,g_{n-1}\,$$\in$$\{0,1\}$\\
$\,$\\
\hline
 $\langle A_{1,x_1} B_{1}^{f_1(g_1)} ...B_{n-1}^{f_{n-1}(g_{n-1})}...A_{n,x_n}\rangle$$=$$\sum\limits_{\mathcal{Y}}(-1)^{h}P_{12^n}(a_1,\vec{b_1},...,\vec{b_{n-1}},...,a_n|x_1,...,x_n)$ \\
  where $\mathcal{Y}$$=$$\{a_1,...,a_n,b_{10},...,b_{1n-1},...,b_{n-10},..., b_{n-1n-1}\}$ \\
  and $h=a_1$$+$$...$$+$$a_n$$+$$s_{f_1(g_1)}(b_{20},...,b_{2n-1})$$+$$...$$+$$s_{f_{n-1}(g_{n-1})}(b_{n-10},...,b_{n-1n-1})$\\\
  $\,$\\
 \hline
 with functions $s_{i-1}(k_1,...,k_n)$ being defined as the sum of all possible product terms of $k_1,...,k_n$ taking $i$ $k_j$'s at a time ($i$$=$$1,...,n-1$). \\
 \hline
\end{tabular}\\
\caption{Detailing of the terms used in Eq.(\ref{n8tt}).}
\end{center}
  	\label{table3}
  \end{table}
  \end{center}
\end{widetext}
Violation of inequalities (Eq.(\ref{n8tt})) for at least one possible $(f_1,...,f_{n-1},g_1,...,g_{n-1})$ ensures non $n$-locality of corresponding correlations. \\
In quantum scenario, let each source generates an $n$-qubit state. Each of the intermediate parties $\mathcal{P}_1^I,...,\mathcal{P}_{n-1}^I$ performs complete $n$ dimensional GHZ basis measurement on the joint of $n$ qubits ($j^{th}$ qubit coming from $\mathcal{\textbf{S}}_i$) whereas each of the extreme $\mathcal{P}_i^E(i$$=$$1,...,n)$ performs projective measurement on its respective qubit. We conjecture that quantum violation of Eq.(\ref{n8tt}) can be obtained. In support of our conjecture we provide a numerical observation for $n$$=$$4,5$. \\
Let each of $n$ independent sources $\mathcal{\textbf{S}}_i$ generates $n$ dimensional GHZ state :
\begin{equation}\label{n20}
\vartheta_n=\frac{|0,0,...0\rangle+|1,1,...,1\rangle}{\sqrt{2}},
\end{equation}
Violation of at least one $n$-local inequality (Eq.(\ref{n8tt})) is obtained. This ensures generation of non $n$-local correlations are generated in the network for $n$$=$$4,5.$\\
\section{Discussions}\label{conc}
In recent past, nonlocality of quantum network correlations under circumstances that some of the parties perform a fixed measurement, has been studied extensively. Topic of our manuscript evolves in this direction.  We analyze nonlocal feature of quantum correlations in networks involving uncorrelated sources when some of the parties do not have the freedom to choose their inputs randomly. Deriving quantum bounds of pre existing \cite{km1} $n$-local inequalities (Eq.(\ref{A3})) turned out to be useful for designing a protocol capable of detecting bipartite resource of entanglement distributed in the network. \\
Analyzing network scenarios involving bipartite entanglement sources, we have then designed networks where sources now generate tripartite quantum states. In this context, we have framed a set of trilocal inequalities (Eq.(\ref{n8t})), violation of which (at least one) is sufficient to guarantee nontrilocality of corresponding correlations. Discussions in Sec\ref{trip}, ensures that randomness in choice of inputs for every party involved is not necessary to generate nonlocal (in sense of nontrilocality) correlations even when tripartite entanglement resources are distributed in the network. Based on numerical evidence we conjecture the same for exploiting non $n$-locality ($n$$\geq$$4$) also. Consequently, even when all the observers cannot randomly select their respective inputs in network scenarios involving $m$-partite ($m$$\geq$$4$) entanglement (generated by sources), nonlocal (non $n$-local) correlations can be obtained. \\
Apart from theoretical perspectives, these trilocal network scenarios turned out to be useful on practical grounds for detection of tripartite entanglement of pure states. More interestingly, protocols designed here can discriminate between some genuinely entangled states and biseparable entanglement existing in any possible grouping of two qubits constituting the three qubit state.
In this context, it will be interesting to enhance the capability of this protocol to discriminate between arbitrary genuine entanglement and biseparable entanglement of any tripartite state. $n$-local non-linear network scenario introduced here may be explored further with an objective to detect entanglement of $m$-partite ($m$$\geq$$4$) states and also to discriminate between genuine entanglement from any other form of $m$-partite entanglement.

\section{Appendix.A}
\begin{widetext}
In \cite{km2}, another trilocal network scenario was introduced where each of the five parties, involved in the network performs one of two dichotomic measurements, i.e., unlike the measurement scenario introduced here, none of the parties has fixed input (for details,see \cite{km2}
).
Correlations generated in such a network \cite{km2} are trilocal if they satisfy:
\begin{equation}\label{n9t}
  \sqrt[3]{|\mathcal{I}_{u_1,u_2,0}|}+\sqrt[3]{|\mathcal{I}_{v_1,v_2,1}|}\leq1\,\,\forall\,u_1,\,u_2,\,v_1,\,v_2\,\in\{0,1\}\,\,\textmd{with,}
\end{equation}
\begin{equation}\label{n9ti}
\mathcal{I}_{u_1(v_1),u_2(v_2),t}=\frac{1}{8}\sum \limits_{x_1,x_2,x_3=0,1}(-1)^{t*q}\langle \mathcal{A}_{1,x_1} \mathcal{B}_{1,y_1=u_1(v_1)} \mathcal{B}_{2,x_2=u_2(v_2)} \mathcal{A}_{2,x_2} \mathcal{A}_{3,x_3}\rangle,~ t\in\{0,1\},\,q=x_1+x_2+x_3
\end{equation}
where
\begin{equation}\label{correl}
    \langle \mathcal{A}_{1,x_1}\mathcal{B}_{1,y_1}\mathcal{B}_{2,y_2} \mathcal{A}_{2,x_2} \mathcal{A}_{3,x_3}\rangle=\sum\limits_{a_1,b_1,b_2,a_2,a_3}(-1)^{m}P(a_1,b_1,b_2,a_2,a_3|x_1,y_1,y_2,x_2,x_3),\, \textmd{with}\,m=a_1+b_1+b_2+a_2+a_3
\end{equation}
 where $x_i$$\in$$\{0,1\}$ denote the input whereas $a_i$$\in$$\{0,1\}$ denote the corresponding output of extreme party $\mathcal{P}_i^{E}(i$$=$$1,2,3).$ Similarly $y_1,y_2$ denote input and $b_1,b_2$ denote output of intermediate party $\mathcal{P}^{I}_1, \mathcal{P}^{I}_2$ respectively. $\mathcal{A}_1, \mathcal{A}_2, \mathcal{A}_3, \mathcal{B}_1, \mathcal{B}_2$ denote the corresponding observables.
We now proceed to prove Theorem.1\\
\textit{Proof:} For simplicity we use the notations $s_{y_i}(b_{i0},b_{i1},b_{i2})$$=$$s_{y_i}$$(i$$=$$1,2)$. Now comparison of the correlation terms related to these two scenarios gives, \\$ P(a_1,b_1,b_2,a_2,a_3|x_1,y_1,y_2,x_2,x_3)=P_{18}(a_1,s_{y_1}=b_1,s_{y_2}=b_2,a_2,a_3|x_1,x_2,x_3)$
\begin{equation}\label{nt3}
 =\qquad\qquad\qquad\sum\limits_{\mathcal{D}}\delta_{b_1,s_{y_1}}
 \delta_{b_2,s_{y_2}}P_{18}(a_1,b_{10},b_{11},b_{12},b_{20},b_{21},b_{22},a_2,a_3|x_1,x_2,x_3),
\end{equation}
$\textmd{where}\,
 \mathcal{D}=\{b_{10},b_{11},b_{12},b_{20},b_{21},b_{22}\}$\\
 $\,$\\
By Eq.(\ref{correl}),\\
$\langle \mathcal{A}_{1,x_1}\mathcal{B}_{1,y_1}\mathcal{B}_{2,y_2} \mathcal{A}_{2,x_2} \mathcal{A}_{3,x_3}\rangle=   \sum\limits_{a_1,a_2,a_3}(-1)^{a_1+a_2+a_3}(P(a_1,0,0,a_2,a_3|x_1,y_1,y_2,x_2,x_3)$$+$$P(a_1,1,1,a_2,a_3|x_1,y_1,y_2,x_2,x_3)$
\begin{equation}\label{stage1}
  - P(a_1,0,1,a_2,a_3|x_1,y_1,y_2,x_2,x_3)-P(a_1,1,0,a_2,a_3|x_1,y_1,y_2,x_2,x_3)).
\end{equation}
Now Eq.(\ref{nt3}) implies, \\ $P(a_1,i,j,a_2,a_3|x_1,y_1,y_2,x_2,x_3)$$=$$\sum\limits_{\mathcal{D}}\delta_{i,s_{y_1}}
 \delta_{j,s_{y_2}}P_{18}(b_{10},b_{11},b_{12},b_{20},b_{21},b_{22},a_2,a_3|x_1,x_2,x_3),\forall\,i,j$$\in$$\{0,1\}.$ \\
Using above relations, in Eq.(\ref{stage1}) and $\mathcal{C}$$=$$\{a_1,a_2,a_3,b_{10}, b_{11},b_{12},b_{20},b_{21},b_{22}\}$ we get:
$$ \langle \mathcal{A}_{1,x_1}\mathcal{B}_{1,y_1}\mathcal{B}_{2,y_2} \mathcal{A}_{2,x_2} \mathcal{A}_{3,x_3}\rangle=   \sum\limits_{\mathcal{C}}(-1)^{a_1+a_2+a_3}\sum\limits_{i,j=0,1}(-1)^{i+j}\delta_{i,s_{y_1}}
 \delta_{j,s_{y_2}}P_{18}(a_1,b_{10},b_{11},b_{12},b_{20},b_{21},b_{22},a_2,a_3|x_1,x_2,x_3)$$
$$\qquad\qquad=\sum\limits_{\mathcal{C}}(-1)^{a_1+a_2+a_3+s_{y_1}+s_{y_2}}P_{18}(a_1, b_{10},b_{11},b_{12},b_{20},b_{21},b_{22},a_2,a_3|x_1,x_2,x_3)$$
Hence,
\begin{equation}\label{stage2}
\langle \mathcal{A}_{1,x_1}\mathcal{B}_{1,y_1}\mathcal{B}_{2,y_2} \mathcal{A}_{2,x_2} \mathcal{A}_{3,x_3}\rangle=\langle A_{1,x_1} B_{1}^{y_1} B_{2}^{y_2}  A_{2,x_2} A_{3,x_3}\rangle
\end{equation}
By Eqs.(\ref{n9t},\ref{n9ti},\ref{correl},\ref{stage2}), we get the required criteria given by Eq.(\ref{n8t}).\\
\end{widetext}
\section{Appendix.B}
\begin{center}
\begin{widetext}
\begin{table}[htp]
\begin{center}
\begin{tabular}{|c|c|c|c|c|c|c|}
\hline
Phase&$\mathcal{P}^E_1$& $\mathcal{P}^E_2$& $\mathcal{P}^E_3$ & $\kappa_1$&$\kappa_2$&$\kappa_3$\\
\hline
$t_{1,1,1}$&$Q_1^{(1)}$&$Q_2^{(1)}$&$Q_3^{(1)}$&$12/3$ or$13/2$ cut&$12/3$ or$13/2$ cut&$12/3$ or$13/2$ cut\\
\hline
$t_{1,1,2}$&$Q_1^{(1)}$&$Q_2^{(1)}$&$Q_3^{(1)}$&$12/3$ or$13/2$ cut&$12/3$ or$13/2$ cut&$12/3$ or$23/1$ cut\\
\hline
$t_{1,1,3}$&$Q_1^{(1)}$&$Q_2^{(1)}$&$Q_3^{(1)}$&$12/3$ or$13/2$ cut&$12/3$ or$13/2$ cut&$23/1$ or$13/2$ cut\\
\hline
$t_{1,2,1}$&$Q_1^{(1)}$&$Q_2^{(2)}$&$Q_3^{(1)}$&$12/3$ or$13/2$ cut&$12/3$ or$23/1$ cut&$12/3$ or$13/2$ cut\\
\hline
$t_{1,2,2}$&$Q_1^{(1)}$&$Q_2^{(2)}$&$Q_3^{(1)}$&$12/3$ or$13/2$ cut&$12/3$ or$23/1$ cut&$12/3$ or$23/1$ cut\\
\hline
$t_{1,2,3}$&$Q_1^{(1)}$&$Q_2^{(2)}$&$Q_3^{(1)}$&$12/3$ or$13/2$ cut&$12/3$ or$23/1$ cut&$23/1$ or$13/2$ cut\\
\hline
$t_{1,3,1}$&$Q_1^{(1)}$&$Q_2^{(3)}$&$Q_3^{(1)}$&$12/3$ or$13/2$ cut&$23/1$ or$13/2$ cut&$12/3$ or$13/2$ cut\\
\hline
$t_{1,3,2}$&$Q_1^{(1)}$&$Q_2^{(3)}$&$Q_3^{(1)}$&$12/3$ or$13/2$ cut&$23/1$ or$13/2$ cut&$12/3$ or$23/1$ cut\\
\hline
$t_{1,3,3}$&$Q_1^{(1)}$&$Q_2^{(3)}$&$Q_3^{(1)}$&$12/3$ or$13/2$ cut&$23/1$ or$13/2$ cut&$23/1$ or$13/2$ cut\\
\hline
$t_{2,1,1}$&$Q_1^{(2)}$&$Q_2^{(1)}$&$Q_3^{(1)}$&$12/3$ or$13/2$ cut&$12/3$ or$13/2$ cut&$12/3$ or$13/2$ cut\\
\hline
$t_{2,1,2}$&$Q_1^{(2)}$&$Q_2^{(1)}$&$Q_3^{(1)}$&$12/3$ or$13/2$ cut&$12/3$ or$13/2$ cut&$12/3$ or$23/1$ cut\\
\hline
$t_{2,1,3}$&$Q_1^{(2)}$&$Q_2^{(1)}$&$Q_3^{(1)}$&$12/3$ or$13/2$ cut&$12/3$ or$13/2$ cut&$23/1$ or$13/2$ cut\\
\hline
$t_{2,2,1}$&$Q_1^{(2)}$&$Q_2^{(2)}$&$Q_3^{(1)}$&$12/3$ or$13/2$ cut&$12/3$ or$23/1$ cut&$12/3$ or$13/2$ cut\\
\hline
$t_{2,2,2}$&$Q_1^{(2)}$&$Q_2^{(2)}$&$Q_3^{(1)}$&$12/3$ or$13/2$ cut&$12/3$ or$23/1$ cut&$12/3$ or$23/1$ cut\\
\hline
$t_{2,2,3}$&$Q_1^{(2)}$&$Q_2^{(2)}$&$Q_3^{(1)}$&$12/3$ or$13/2$ cut&$12/3$ or$23/1$ cut&$23/1$ or$13/2$ cut\\
\hline
$t_{2,3,1}$&$Q_1^{(2)}$&$Q_2^{(3)}$&$Q_3^{(1)}$&$12/3$ or$13/2$ cut&$23/1$ or$13/2$ cut&$12/3$ or$13/2$ cut\\
\hline
$t_{2,3,2}$&$Q_1^{(2)}$&$Q_2^{(3)}$&$Q_3^{(1)}$&$12/3$ or$13/2$ cut&$23/1$ or$13/2$ cut&$12/3$ or$23/1$ cut\\
\hline
$t_{2,3,3}$&$Q_1^{(2)}$&$Q_2^{(3)}$&$Q_3^{(1)}$&$12/3$ or$13/2$ cut&$23/1$ or$13/2$ cut&$23/1$ or$13/2$ cut\\
\hline
$t_{3,1,1}$&$Q_1^{(3)}$&$Q_2^{(1)}$&$Q_3^{(1)}$&$12/3$ or$13/2$ cut&$12/3$ or$13/2$ cut&$12/3$ or$13/2$ cut\\
\hline
$t_{3,1,2}$&$Q_1^{(3)}$&$Q_2^{(1)}$&$Q_3^{(1)}$&$12/3$ or$13/2$ cut&$12/3$ or$13/2$ cut&$12/3$ or$23/1$ cut\\
\hline
$t_{3,1,3}$&$Q_1^{(3)}$&$Q_2^{(1)}$&$Q_3^{(1)}$&$12/3$ or$13/2$ cut&$12/3$ or$13/2$ cut&$23/1$ or$13/2$ cut\\
\hline
$t_{3,2,1}$&$Q_1^{(3)}$&$Q_2^{(2)}$&$Q_3^{(1)}$&$12/3$ or$13/2$ cut&$12/3$ or$23/1$ cut&$12/3$ or$13/2$ cut\\
\hline
$t_{3,2,2}$&$Q_1^{(3)}$&$Q_2^{(2)}$&$Q_3^{(1)}$&$12/3$ or$13/2$ cut&$12/3$ or$23/1$ cut&$12/3$ or$23/1$ cut\\
\hline
$t_{3,2,3}$&$Q_1^{(3)}$&$Q_2^{(2)}$&$Q_3^{(1)}$&$12/3$ or$13/2$ cut&$12/3$ or$23/1$ cut&$23/1$ or$13/2$ cut\\
\hline
$t_{3,3,1}$&$Q_1^{(3)}$&$Q_2^{(3)}$&$Q_3^{(1)}$&$12/3$ or$13/2$ cut&$23/1$ or$13/2$ cut&$12/3$ or$13/2$ cut\\
\hline
$t_{3,3,2}$&$Q_1^{(3)}$&$Q_2^{(3)}$&$Q_3^{(1)}$&$12/3$ or$13/2$ cut&$23/1$ or$13/2$ cut&$12/3$ or$23/1$ cut\\
\hline
$t_{3,3,3}$&$Q_1^{(3)}$&$Q_2^{(3)}$&$Q_3^{(1)}$&$12/3$ or$13/2$ cut&$23/1$ or$13/2$ cut&$23/1$ or$13/2$ cut\\
\hline
\end{tabular}\\
\caption{Detailed distribution of qubits among the extreme parties in phases of the protocol. $\forall j,k$$\in$$\{1,2,3\},$ $Q_j^{k}$ denotes $k^{th}$ qubit of $\kappa_j.$ For each $i$$=$$1,2,3,$ $(i+4)^{th}$ column of the table denotes possible nature of entanglement of unknown state $\kappa_i$ other than genuine entanglement when violation of at least one trilocal inequality is obtained in the corresponding phase.}
\end{center}
  	\label{table3}
  \end{table}
\end{widetext}
  \end{center}
As already mentioned in the main text that distribution of qubits among the extreme parties is crucial in the context of obtaining violation by biseparable entanglement. So for designing the protocol for purpose of detecting tripartite entanglement all possible arrangement of qubits among the extreme parties are considered. At this junction, one may recall that in the non-linear trilocal network scenario (Fig.3), for a fixed source, pattern of arranging qubits among the intermediate parties does not contribute in detecting nature of biseparable entanglement. So distribution of qubits only among the extreme parties $\mathcal{P}_1^E,$ $\mathcal{P}_2^E$ and $\mathcal{P}_3^E$ is enlisted in Table.VI. The last three columns of Table.VI indicate the possible nature of biseparable entanglement of the unknown state under the circumstance that violation of at least one trilocal inequalities (Eq.(\ref{n8t})) is obtained in the corresponding phase. For instance, consider the phase $t_{1,2,3}.$ If violation is obtained in this phase of the protocol, then following are the possible nature of the three unknown quantum states:
\begin{itemize}
\item $\kappa_1$ is either genuinely entangled or have biseparable entanglement content in $12/3$ or $13/2$ cut.\\
\item$\kappa_2$ is either genuinely entangled or have biseparable entanglement content in $12/3$ or $23/1$ cut.\\
\item$\kappa_3$ is either genuinely entangled or have biseparable entanglement content in $23/1$ or $13/2$ cut.\\
\end{itemize}

\begin{thebibliography}{1}
\bibitem{horodecki11} Horodecki, R., Horodecki, P., Horodecki, M. and  Horodecki, K. : ``\textit{Quantum Entanglement}'', Rev. Mod. Phys. \textbf{81}, 865 (2009).
\bibitem{RRR} Bell, J. S. : ``\textit{On the Einstein Podolsky Rosen Paradox}'', Physics \textbf{1}, 195 (1964).
\bibitem{RLL} Bell, J. S. : ``\textit{Speakable and Unspeakable in Quantum Mechanics}'', Cambridge University Press England (1987).
\bibitem{as1} Groblacher, S.\emph{et al.} : ``\textit{ An experimental test of non-local realism.}'' Nature \textbf{446}, 871-875 (2007).
 \bibitem{as} Aspect, A., Dalibard, J. Roger, G. : ``\textit{ Experimental test of Bell's inequalities using time-varying analyzers.}'', Phys. Rev. Lett.  \textbf{49}, 1804-1807 (1982).
\bibitem{Cl} Clauser, J.F., Horne, M.A., Shimony, A. and Holt, R.A. : ``\textit{ Proposed experiment to test local hidden-variable theories}'', Phys. Rev.Lett. \textbf{23}, 880 (1969).
\bibitem{ng1}Gisin, N. \emph{et al.} : ``\textit{All entangled pure quantum states violate the bilocality inequality.}'', Phys. Rev. A \textbf{96}, 020304 (2017).
\bibitem{ng2}Gisin, N. : ``\textit{The Elegant Joint Quantum Measurement and some conjectures about N-locality in the Triangle and other Configurations}'', arXiv:\textbf{1708.05556} [quant-ph].
\bibitem{ng3}Renou, M. R. \emph{et al.} : ``\textit{Genuine quantum nonlocality in the triangle network}''  arXiv:\textbf{1905.04902}v2[quant-ph].
\bibitem{ng4} T. Fraser, E. Wolfe, : ``\textit{Causal compatibility inequalities admitting quantum violations in the triangle structure}'', Phys. Rev. A \textbf{98}, 022113 (2018).
\bibitem{ng5}Renou, M. R. \emph{et al.} : ``\textit{Limits on correlations in networks for quantum and no-signaling resources}'', Phys. Rev. Lett. \textbf{ 123}, 070403 (2019).
\bibitem{ng6} N. Gisin, Entropy \textbf{21}, 325 (2019).
\bibitem{BRAN} C. Branciard,D. Rosset, N. Gisin and S. Pironio, Phys. Rev. A \textbf{85}, 032119 (2012).
\bibitem{BRA} C. Branciard, N. Gisin, and S. Pironio, Phys. Rev. Lett. \textbf{104},170401 (2010).
\bibitem{star} A. Tavakoli, P. Skrzypczyk, D. Cavalcanti and A. Ac\'{i}n : ``\textit{Nonlocal correlations in the star-network configuration}'',Phys. Rev. A \textbf{90}, 062109.
\bibitem{km1} K. Mukherjee, B. Paul and D. Sarkar: ``\textit{Correlations In n-local Scenario}'' Quantum Inf Process. \textbf{14}, 2025 (2015).
\bibitem{km2} K. Mukherjee, B. Paul and D. Sarkar:``\textit{Nontrilocality: Exploiting nonlocality from three-particle systems}'', Phys. Rev. A \textbf{96}, 022103 (2017).
\bibitem{km3} K. Mukherjee, B. Paul and D. Sarkar:``\textit{Revealing advantage in a quantum network}'', Quantum Inf Process. \textbf{15}, 7,2895-2921 (2016).
\bibitem{km4} K. Mukherjee, B. Paul and D. Sarkar:``\textit{Restricted distribution of quantum correlations in bilocal network}'', Quantum Inf Process \textbf{18}, 212 (2019).
\bibitem{key1} J. Barrett, L. Hardy, and A. Kent, Phys. Rev. Lett. \textbf{ 95}, 010503 (2005).
\bibitem{Mayer} D. Mayers and A. Yao,\,in Proceedings of the 39th IEEE Symposiumon Foundations of Computer Science (IEEE Computer Society, Los Alamitos  CA, USA,1998)p.503.
\bibitem{Acin}A. Ac\'{i}n, N. Brunner, N. Gisin, S. Massar, S. Pironio, and V. Scarani, Phys. Rev. Lett. \textbf{98}, 230501 (2007).
\bibitem{key2} A. Ac\'{i}n, N. Gisin, and L. Masanes, Phys. Rev. Lett. \textbf{97}, 120405 (2006).
\bibitem{Colbeck} R. Colbeck and A. Kent, Journal of Physics A: Mathematical and Theoretical \textbf{44}, 095305 (2011).
\bibitem{Pironio} S. Pironio, A. Ac\'{i}n, S. Massar, A. B. de la Giroday, D. N. Matsukevich,P. Maunz, S. Olmschenk, D. Hayes, L. Luo, T. A.Manning, and C. Monroe, Nature \textbf{464}, 1021 (2010).
 \bibitem{game}    N. Brunner and N. Linden, Nature Communications \textbf{ 4}, 2057 (2013).
 \bibitem{preskil} Preskill, J., ``\textit{Lecture Notes for Physics: Quantum Information and
Computation}'', (California Institute of Technology) (1998).
\bibitem{HOR} Horodecki, R., Horodecki, P. and Horodecki, M. : ``\textit{Violating Bell inequality by mixed spin$-\frac{1}{2}$ states: necessary and sufficient condition}'', Phys. Lett. A \textbf{200}, 340 (1995).
\bibitem{ACN} K. Mukherjee, B. Paul, D. Sarkar : ``\textit{Efficient test to demonstrate genuine three particle nonlocality}'', J. Phys. A: Math. Theor. \textbf{48}, 2025 (2015).
\bibitem{ACN5} Ac\'{i}n, A. : ``\textit{Generalized Schmidt Decomposition and Classification of Three-Quantum-Bit States}'', Phys. Rev. Lett. \textbf{85}, 1560 (2000).
\end{thebibliography}
\end{document}